\documentclass[fleqn,usenatbib]{mnras}

\usepackage{newtxtext,newtxmath}

\usepackage[T1]{fontenc}
\usepackage[utf8]{inputenc}
\usepackage{ae,aecompl}

\usepackage{graphicx}	
\usepackage{amsmath}	
\usepackage{amssymb}	

\title[GCs in galaxy clusters]{Simulating the Spatial
  Distribution and Kinematics of Globular Clusters within Galaxy
  Clusters in Illustris}

\author[F. Ramos-Almendares et al.]{
\parbox[t]{\textwidth}{
Felipe Ramos-Almendares$^{1}$\thanks{E-mail: framos@oac.unc.edu.ar},
Laura V. Sales$^{2}$\thanks{Hellman Fellow},
Mario G. Abadi$^{1,3}$,
Jessica E. Doppel$^{2}$,
Hernan Muriel$^{1,3}$,
and Eric W. Peng$^{4,5}$
}
\\
\\
$^{1}$CONICET-Universidad Nacional de C\'ordoba, Instituto de Astronomía Te\'orica y Experimental (IATE), C\'ordoba, Argentina\\
$^{2}$University of California Riverside, 900 University Ave., Riverside CA 92521, USA\\
$^{3}$Observatorio Astron\'omico, Universidad Nacional de C\'ordoba,  C\'ordoba, Argentina\\
$^{4}$Department of Astronomy, Peking University, 5 Yiheyuan Road, Beijing, China 100871\\
$^{5}$Kavli Institute for Astronomy and Astrophysics, Peking University, 5 Yiheyuan Road, Beijing, China 100871\\
}

\pubyear{2019}

\begin{document}
\label{firstpage}
\pagerange{\pageref{firstpage}--\pageref{lastpage}}
\maketitle

\begin{abstract}
  We study the assembly of globular clusters (GCs) in
  $9$ galaxy clusters using the cosmological simulation
  Illustris.   GCs are tagged to individual galaxies at their infall time. The tidal removal of GCs from their galaxies and the distribution of the GCs within the cluster is later followed self-consistently by the simulation. 
  The method relies on the simple assumption of a single power-law 
  relation between halo mass ($M_{\rm vir}$) and mass in GCs ($M_{\rm GC}$)
  as found in observations. We find that the GCs
  specific frequency $S_N$ as a function of $V$-band magnitude naturally
  reproduces the observed ``U''-shape due to the 
  combination of the power law $M_{GC}$-$M_{\rm vir}$ relation and the 
  non-linear stellar mass ($M_*$) - halo mass relation from the simulation. 
  Additional scatter in the $S_N$ values is traced back to 
  galaxies with early infall times due to the evolution of the 
  $M_*$-$M_{\rm vir}$ relation with redshift.
  GCs that have been tidally removed from their galaxies form the present-day
  intra-cluster component, from which about $\sim 60\%$ were brought
  in by galaxies that currently orbit within the cluster potential. The
  remaining ``orphan'' GCs are contributed by satellite galaxies
  with a wide range of stellar masses that are fully tidally disrupted
  at $z=0$. This intra-cluster component is 
  a good dynamical tracer of the dark matter potential. 
  As a consequence of the accreted nature of most
  intra-cluster GCs, their orbits are fairly radial with a predicted
  orbital anisotropy $\beta \geq 0.5$. However, local tangential motions 
  may appear as a consequence of localized substructure, providing a 
  possible interpretation to the $\beta<0$ values suggested in observations 
  of M87.
\end{abstract}

\begin{keywords}
galaxies: clusters: general -- galaxies: haloes -- galaxies: star clusters: general
\end{keywords}



\section{Introduction}
Galaxy clusters are populated by tens of thousands of globular clusters (GCs)
that distribute around galaxies as well as in intra-cluster space. 
However, little is known about their assembly history and their connection to the
cluster build up within the cosmological framework. GCs are among the 
densest stellar systems in the Universe, with typical stellar masses in 
the range $[10^4$-$10^6\; \rm M_\odot]$ and sizes of only a few parsecs \citep{harris_globular_1979,brodie_extragalactic_2006}.
 Because of their inferred old ages \citep{vandenberg_age_1996} and 
clustered distribution around galaxies, they are believed to be surviving 
probes of the star formation in the early universe.

GCs are, however, more metal poor than the bulk of the stars in
their host galaxy. In fact, GCs display a wide
range of colors and metallicities that suggests different origins.
One possible interpretation is that the youngest and more metal rich GCs are formed {\it in-situ}
while the metal poor component is likely acquired hierarchically via
the accretion of smaller galaxies
\newpage
\citep[e.g., ][]{harris_globular_1991, ashman_formation_1992, zepf_globular_1993,cote_formation_1998,kissler-patig_constraints_1998,gebhardt_globular_1999}.
This is consistent with the exquisite GCs data around the Milky Way 
(MW), other galaxies in the Local Group \citep{harris1996,helmi2018,johnson2015} and in massive ellipticals
\citep{forbes_evidence_2011,usher_sluggs_2012,taylor_survey_2017}. In terms of numbers, however, the rich 
environment of nearby galaxy clusters, due to their high matter density and 
wide range of galaxy masses and morphologies, offer
the best opportunity to understand the connection between
GCs, galaxies and ultimately their dark matter halos.

Significant effort and resources in the community
have focused on the generation of comprehensive GC maps in nearby galaxy clusters 
such as Virgo \citep{jordan_acs_2009,durrell_next_2014}, Fornax \citep{liu2019} 
and Coma \citep{madrid2018}.
Along with other studies, these data have raised a number of interesting discoveries that 
shed light on the formation of GCs. First, the relation between the total mass in 
GCs ($M_{\rm GC}$) and the estimated dark matter halo mass ($M_{\rm vir}$) for galaxies
follows a single and relatively tight power-law correlation \citep{harris_dark_2015,hudson_dark_2014,harris_catalog_2013,blakeslee_globular_1997,peng_acs_2008,georgiev_globular_2010,spitler_new_2009,forbes2018} that contrasts the complex and non-linear 
correlation of galaxy stellar mass ($M_*$) and halo mass suggested by abundance matching models 
\citep{moster_galactic_2013, behroozi_average_2013}. 
Second, the specific frequency of GCs ($S_N$),
defined as the number of GCs normalized to the galaxy luminosity, is also highly non-linear, 
with dwarf galaxies in clusters that show high $S_N \geq 5$ values which are at odds with
dwarf galaxies of similar mass located in the field and that are characterized by $S_N \sim 2$ \citep{peng_acs_2008,georgiev_globular_2010}. 

Some of the observed trends in GCs can be explained in models where they trace the most turbulent, high
density and gas rich star formation episodes in galaxies \citep{Kravtsov2005,prieto_dynamical_2008},
and analytical calculations support this view \citep{kruijssen_initial_2012, kruijssen_globular_2015,elmegreen_globular_2017}. 
Other scenarios where GCs are placed at the centers of their own dark matter halo 
\citep{peebles_dark_1984,rosenblatt_pregalactic_1988}
and form completely independent of their host galaxy are, although compelling, currently disfavoured 
due to the large degree of tidal stripping expected 
and the observationally constrained abundances and radial distribution of GCs around galaxies  \citep{carlberg_globular_2018,creasey_globular_2019}. 

While other mechanisms
such as baryonic flows in the early universe may suggest the possible formation of 
GC-like objects \citep{naoz_globular_2014,chiou2019}, 
the  most dominant formation channel for GCs seems therefore to be related 
to galaxies undergoing high-density and violent star formation activity. 
Additional observational support in this direction comes from the observed 
correlation between star formation density and an increased fraction
of stars born in bound stellar clusters in nearby star forming galaxies
\cite{goddard2010, adamo2011, bastian_constraining_2013}. 

Due to the low mass and small sizes of GCs, resolving these
systems in hydrodynamical simulations of galaxy formation is a daunting task, 
especially when embedded within the cosmological framework. 
  Encouragingly, full cosmological zoom-in simulations are starting to resolve
 the formation of individual GC-like objects \citep{kim_formation_2018,Ma2019,Lahen2019}. 
 However, due to the expensive calculations the runs are necessarily stopped at
high redshift ($z=5$) or they focus on very low mass dwarfs, making them 
not fully suitable for predictions of GCs in nearby galaxies and groups in the local Universe. 
Several theoretical efforts are currently underway
that couple the formation and evolution of stellar clusters to the star particles formed in cosmological
hydrodynamical simulations \citep{li_star_2017,renaud_origin_2017, li_star_2018,
pfeffer_e-mosaics_2018,li_star_2019}. However, the price to pay for such an approach
is high, as it requires resolving 
the necessary conditions for molecular clouds formation and their posterior
tidal evolution at the sub-pc level. Currently this is only possible at the scale of
individual $\sim L_*$ and dwarf galaxies, but not for larger systems
such as groups and clusters. To take advantage of the richest set of observational
data of GCs available to date, which is collected from all nearby galaxy clusters, 
one must therefore look for alternative and numerically more efficient techniques. 

One possibility is to study the abundance and chemical properties of GCs by following 
their formation, evolution and destruction via semi-analytical modeling built on top of
cosmological N-body simulations \citep{el-badry_formation_2019,choksi_formation_2019,Muratov2010,Tonini2013,li_modeling_2014,boylan-kolchin2017,Choksi2018}. 
Such models are powerful since they make quick predictions over a wide range
of halo masses and assembly histories by simultaneously sampling the large parameter
space that is inherent to following the formation of GCs. But it is 
also desirable to make predictions on the spatial distribution or kinematics
of these simulated GCs. A possible avenue to achieve the latter is by 
implementing {\it tagging}
techniques, where GCs are ``painted'' on top of other particle types in the simulation,
i.e. dark matter or stars, and use their dynamical evolution to self-consistently follow
the phase space information expected for the GCs within the hierarchical assembly of
structures in $\Lambda$CDM.
Such an approach, reminiscent of the insightful \citet{bullock_tracing_2005} technique
used in stellar halo analysis, has been implemented to study the stripping of GCs 
within clusters \citep{bekki_dynamical_2003, ramos_tidal_2015, ramos-almendares_intra-cluster_2018}
and the large specific frequency of dwarf ellipticals in clusters \citep{mistani_assembly_2016}. 

In this work, we apply a similar technique to that presented in \citet{ramos-almendares_intra-cluster_2018},
but instead of using  a single N-body only zoom-in simulation of a galaxy cluster, we couple it to
$9$ galaxy clusters selected from the hydrodynamical cosmological simulation Illustris \citep{vogelsberger2014a, vogelsberger2014b}. 
 For the first time, this allows us to make predictions on the connection between GCs
 and the properties of {\it galaxies}, for which the stellar mass build up and gas evolution
 is naturally followed by the hydrodynamical treatment
 in the simulations.
 Our work is different from that presented in \citep{mistani_assembly_2016}
 (which also used Illustris) in that the abundances of GCs are calculated from the
 dark matter halo mass by calibrating
 the model to the observed $M_{\rm GC}$-$M_{\rm vir}$ relation presented in \citet{harris_dark_2015}
 instead of modeling the GC formation by using the individual star formation histories of 
 galaxies as given by the simulation. Furthermore, our model extends that of 
 \citet{mistani_assembly_2016} in that all galaxies ever entering the galaxy clusters 
 are tagged with GCs rather than looking only at surviving dwarf galaxies
 and
 that the red/blue GC colors are assigned following observational fractions of \citet{harris_dark_2015}. 
 
 Our technique delivers GC maps in galaxy clusters that can be directly compared
 to observations in Virgo and the Fornax cluster. In this paper, the first of a series, 
 we will introduce the technique (Sec.~\ref{sec:method}) and study the resulting specific frequency
 $S_N$ of galaxies (Sec.\ref{sec:s_n}) as well as the predictions for the radial distribution
 and kinematics of the intra-cluster GC component (Sec.~\ref{sec:rad_vel}). We conclude by summarizing
 our main results in  Sec.~\ref{sec:concl}.

\begin{figure*}
	\includegraphics[width=\textwidth]{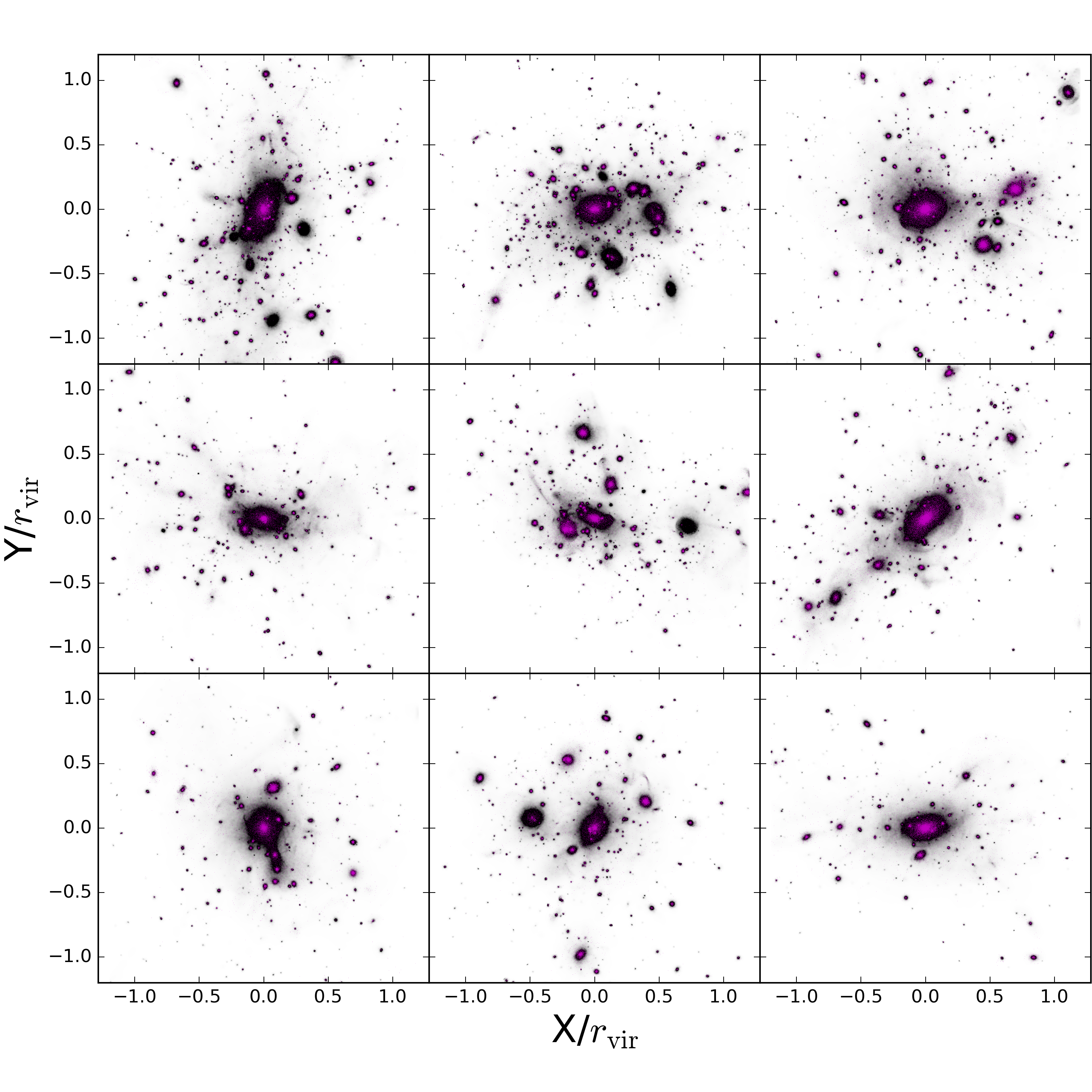}
    \caption{Projected map of our simulated clusters in Illustris selected with 
    $M_{\rm vir} \sim 10^{14}\; \rm M_\odot$. Gray scale shows the stellar density distribution
    and in magenta we highlight the generated catalog of GCs. Note that GCs tend
    to cluster around galaxies, but are also found at present day in an extended 
    and diffuse intra-cluster component (see also Fig. \ref{fig:zoom-in} for a zoom-in of the top-right panel cluster).
    }
    \label{fig:map}
\end{figure*}
\section{Methods}
\label{sec:method}

The Illustris suite of simulations consists of a series of 
cosmological boxes $106$ Mpc on one side that follow
the assembly of tens of thousands of halos and galaxies from $z=127$ until present day
\citep{vogelsberger2014a, vogelsberger2014b, genel2014}. The simulation adopts a cosmology consistent
 with the  Wilkinson Microwave Anisotropy Probe (WMAP)-9 measurements 
\citep{hinshaw2013} and is run using 
the {\sc arepo} hydrodynamical code \citep{springel2010}. Cooling, star formation, 
stellar feedback and stellar evolution are modeled following the implementations described 
in \citet{vogelsberger2013}, while black hole growth and feedback is modeled following
\citet{sijacki2015}. Relevant to the science presented in this paper, Illustris
has been shown to reproduce several of the observables at the galaxy population
level, including the scaling relations and angular momentum content \citep{genel2019}, 
the estimated merger rates \citep{rodgomez2015}, general morphology
diversity \citep{snyder2015, rodgomez2017} and color evolution of satellite
galaxies \citep{sales2015}, among others. 

The Illustris suite is composed of three different resolution levels and sibling
runs with/without the inclusion of baryons. We use for our study the highest 
resolution run (Illustris-1 referred to as Illustris hereafter for simplicity) 
achieving a mass per particle
$m_{\rm p,\rm g} = 1.3 \times 10^6 \; \rm M_\odot$ and 
$m_{\rm p,\rm dm} = 6.3 \times 10^6 \; \rm M_\odot$ for gas and dark matter, 
respectively, and a gravitational softening that is kept always smaller than $700$ pc
for the collisionless components (dark and stars). Note that the high resolution 
gas typically resolves much smaller scales due to the adaptive Voronoi mesh construction,
reaching a typical cell size $\sim 50$ pc at the centers of galaxies. 

We use galaxy and halo catalogs from {\sc subfind} \citep{springel2001, dolag2009}
which run over pre-identified Friends-of-Friends (FoF) groups \citep{davis1985}. We
will use the commonly accepted terminology of ``centrals" vs. ``satellites" to refer either
to the main galaxy siting at the center of the gravitational potential (a.k.a {\it central})
or to the rest of the galaxies associated to the same FoF group ({\it satellites}).
Further details on the simulations data products can be found in \citet{nelson2015}. 

\subsection{Galaxy sample and infall time}

We analyze $9$ galaxy clusters from Illustris selected to have present day virial mass
$M_{\rm vir} \geq 8 \times 10^{13}\; \rm M_\odot$. In what follows, we use $200$
times the critical density of the universe to define all virial quantities. 
Although $10$ objects satisfy our criteria at $z=0$, group 5 is not included
in our sample due to problems in tracing the central galaxy of the group backwards
in time. Our sample then comprises $9$ galaxy clusters with virial masses consistent with
those of Virgo and Fornax. 

We use the {\sc SubLink} merger trees \citep{rodgomez2015} to follow all
galaxies that ever interacted with each of our clusters. For GC tagging, we choose
all galaxies with stellar mass above $M_* =10^8 \; \rm M_\odot$ at infall time, 
ensuring that our galaxies are numerically well resolved ($\geq 60$ stellar particles). 
We define the infall time as the last snapshot where a galaxy is a central of its own
FoF group. We note that such a definition does not necessarily correspond to the time
when they join the cluster but can instead be earlier if a galaxy joins a different
group before infalling onto the cluster. However, choosing a different infall time
will not impact our results significantly, as the dark matter halo of any galaxy
(which in our model is what determines the amount and distribution of GCs) stops
growing after they become satellites of any system.
 The central galaxy of each cluster requires a different treatment as their 
infall times are not well defined. 
We choose to tag them at the time when they reach $5\%$ of their final virial mass 
in an attempt to capture their individual GC population separated from that resulting from the later hierarchical
assembly of the clusters. With this definition, we find $z\sim 2.5$-$3.0$ as the 
typical tagging redshifts for our centrals. We have explicitly checked that our 
results are not sensitive to this particular choice.
On average, our selection
criteria results in $\sim 300$ galaxies with GCs tagged per galaxy cluster.

\subsection{GC Tagging}
   
  In order to select dark matter particles as tracers of globular cluster systems we use the  technique outlined by \citet{bullock_tracing_2005} and \citet{penarrubia_tidal_2008}. 
  This technique selects a subsample of particles that follow a chosen spatial density profile, and it has been used by the above authors to simulate stellar populations in dark matter only simulations.
  This method has been already successfully implemented to model the GC systems of galaxies in dark matter only simulations by our team  \citep{ramos_tidal_2015,ramos-almendares_intra-cluster_2018}. Briefly, the procedure is as follows.

 First we find the best Navarro-Frenk-White \citet[][NFW]{navarro_structure_1996} fit that 
 describes the density profile of the dark matter in a given infalling halo:

\begin{equation}
\rho_{NFW}(r)=\frac{\rho^0_{NFW}}{(r/r_{NFW})(1+r/r_{NFW})^2} 
\end{equation}
 where $\rho^0_{NFW}$ and $r_{NFW}$ are the dark matter halo characteristic density and scale length, respectively. We assume that $r_{NFW}$ can be approximated by $r_{max}=\alpha r_{NFW}$ \citep{bullock_profiles_2001}, where $\alpha=2.1623$ and $r_{max}$ is the radius of the maximum circular velocity, a parameter provided by the {\sc subfind} catalog in the Illustris database.

 As in \citet{ramos_tidal_2015} and \citet{ramos-almendares_intra-cluster_2018}, we choose the density profiles of globular cluster systems to follow a \citet{hernquist_analytical_1990} profile:
\begin{equation}
\rho_{HQ}(r)=\frac{\rho^0_{HQ}}{(r/r_{HQ})(1+r/r_{HQ})^3} 
\label{eq:rho}
\end{equation}
where  $\rho^0_{HQ}$ and $r_{HQ}$ are the \citep{hernquist_analytical_1990} characteristic density and scale length, respectively. 

 The radial distribution of GCs around galaxies have been studied in observations \citep{dirsch2003,dirsch2003b,bassino_large-scale_2006,coenda_tidal_2009} and shown to be spatially more extended than the stars but less extended that the inferred underlying dark matter halo \citep{Abadi2006,hudson2018}. Furthermore, for the Milky Way galaxy there seems to be a {\it dichotomy} of GC properties: 
a more centrally concentrated distribution of red, younger and more metal rich GCs, and a more extended old and metal poor population of ``blue" GCs with presumably accreted origins. Interestingly, this dichotomy or bi-modality may not hold true in other external galaxies, which instead may be better explained by a {\it continuous gradient} on their GC population \citep{harris_galactic_2017,forte2017,forte2019}. 

Regardless of this dichotomy vs. gradient discussion -- which is in itself worth of in-depth theoretical investigations that is beyond the scope of this paper--  observational evidence suggests that the redder GC component tends to be more centrally clustered compared to the bluer GCs. In this work, we attempt to model this spatial segregation by tagging two different GC populations, ``blue'' and ``red", with different radial distributions. Notice, however, that this does not require the GC systems to be bi-modal, but instead can be interpreted as a rough description of the ``redder" and ``bluer" parts of a continuous distribution.

In Eq.~\ref{eq:rho} take $r_{HQ}=\beta r_{NFW}$, with $\beta=3.0$ and $\beta=0.5$ for blue and red globular clusters, respectively.  The specific values for $\beta$ were obtained as least square fits to the projected density profiles of red and blue GCs in observations at $z=0$ \citep[for details please see ][]{ramos_tidal_2015}.

 Once we have the right scale lengths for these density profiles, we compute the distribution function for dark matter ($f_{NFW}$) and globular cluster systems ($f_{HQ}$) using the equation

\begin{equation}
f(\epsilon)=\frac{1}{8\pi}
\Big[\int_0^\epsilon\frac{d^2\rho}{d\psi^2}\frac{d\psi}{\sqrt{\epsilon-\psi}}+\frac{1}{\sqrt{\epsilon}}\Big(\frac{d\rho}{d\psi}\Big)_{\psi=0}\Big]
\label{ec:fdist}
\end{equation}

\noindent
where $\psi$ and $\epsilon$ are the relative gravitational potential and the relative energy, respectively \citep{binney_galactic_1987}.

This leaves us in the position to select, in bins of relative energy $(\epsilon,\Delta \epsilon)$, a fraction $f_{HQ}/f_{NFW}$ of particles as candidate GC particles.

To guarantee that the GC system is more concentrated than the dark matter halo as inferred from observations, 
we truncate the selection to a radius $r_{CutOff}=r_{50,h}/3$, where $r_{50,h}$ is the half mass radius of the dark matter halo.  This choice has been used in previous work and it is inspired on the typical radius for the Milky Way GCs \citep{yahagi_formation_2005,ramos_tidal_2015,ramos-almendares_intra-cluster_2018}. 

\begin{figure}
	\includegraphics[width=\columnwidth]{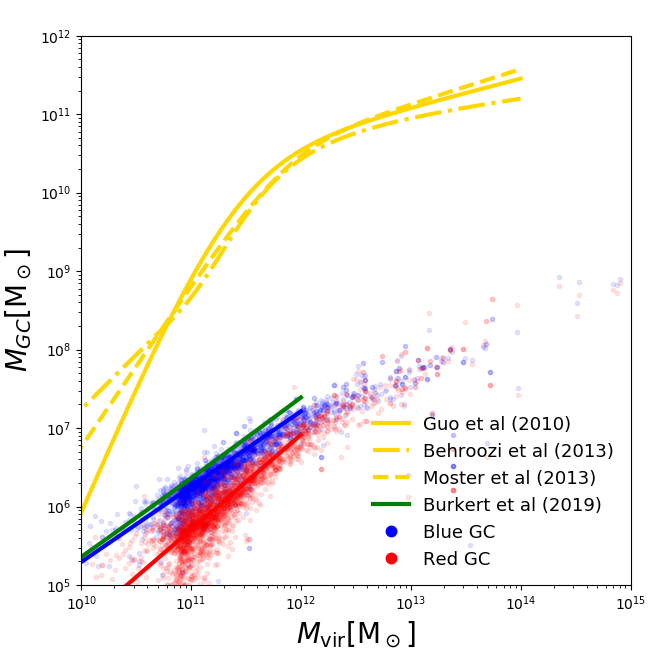}
    \caption{Mass in GCs at the present day in our galaxies as a function of their halo mass. 
    The method is calibrated to reproduce the \citet{harris_dark_2015} measurements for the
    red and blue components (solid lines). Green solid line shows the combined mass from both
    populations taken from the compilation in \citet{burkert2019}. For comparison, we also 
    display in yellow lines the relation between
    halo mass and stellar mass in galaxies following the abundance matching 
    models presented in \citet{guo_how_2010} 
    (solid), \citet{moster_galactic_2013} (dashed) and \citet{behroozi_average_2013} (dotted). 
    The stellar mass - halo mass relation is
    markedly different from the power law found in the GCs mass relation to halo mass, a fact that might
    drive the characteristic ``U"-shape in the 
    specific frequency of galaxies as a function of their luminosity (see Fig.~\ref{fig:s_n}).}
    \label{fig:fig2}
\end{figure}

\subsection{GC Mass}
\label{ssec:gcmass}
In order to assign masses to particles tagged as GCs, we use \citet{harris_dark_2015} as observational constraints.
These authors found that, at z=0, the total mass of globular clusters inside a galaxy can be described by

\begin{equation}
M_{\rm GC,0} = a \; M_{\rm vir,0}^b
\label{eq:harris}
\end{equation}
 
\noindent 
where $M_{\rm GC,0}$ and $M_{\rm vir,0}$ are the mass of GCs and the halo mass of the galaxy at $z=0$ in solar mass units ($M_\odot$), respectively, and taking $(a,b)= (4.9 \times 10^{-5}, 0.96)$ for blue and $(a,b)=(2.6 \times 10^{-8},1.2)$ for red GCs. \citet{harris_dark_2015} estimate $M_{\rm vir,0}$ from the stellar mass of the galaxies using the empirical star-halo mass relation from lensing results by \citet{hudson_cfhtlens:_2015} (see their Appendix C, eq. C1).

We use this result to compute the GC mass at infall in our simulated galaxies using:
\begin{equation}
M_{\rm GC,\rm inf} = a_{\rm inf} \;  M_{\rm vir,\rm inf}^{b_{\rm inf}}
\label{eq:harriszin}
\end{equation}

\noindent
where $M_{\rm GC,\rm inf}$ and $M_{\rm vir,\rm inf}$ are the mass of GCs and the halo mass of the 
galaxy at infall time in solar mass units, respectively. 
Since our definition of infall time ensures that the galaxy 
is a central galaxy of a FoF group at this time, we take $M_{\rm vir,\rm inf} = M_{\rm vir}$.

The values of $a_{\rm inf}$ and $b_{\rm inf}$ in Eq.~\ref{eq:harriszin} need to be {\it calibrated} so that, 
at redshift $z=0$ the resulting correlation between $M_{\rm vir,0}$ and $M_{\rm GC,0}$ reproduces
the observed normalization from \citet{harris_dark_2015}. This exercise gives us
$a_{\rm inf}=3.5\times10^{-4} \; (2.0\times10^{-7})$ and $b_{\rm inf}=0.9 \; (1.15)$ for blue (red) GC, respectively.  This calibration is needed to account for the average fraction of GCs lost by 
each galaxy to tidal stripping in the cluster potential.
We note that because at $z=0$ the
simulated galaxies are satellites within the cluster (and therefore their halo/virial
mass is ill defined) to make a fair comparison to \citet{harris_dark_2015}, we compute
the $z=0$ halo mass using the same procedure as the authors and implement the 
star-mass relation from \citet{hudson_cfhtlens:_2015}
using the stellar mass of the galaxies at $z=0$ as provided by the simulation. 

\begin{figure} 
\includegraphics[width=\columnwidth]{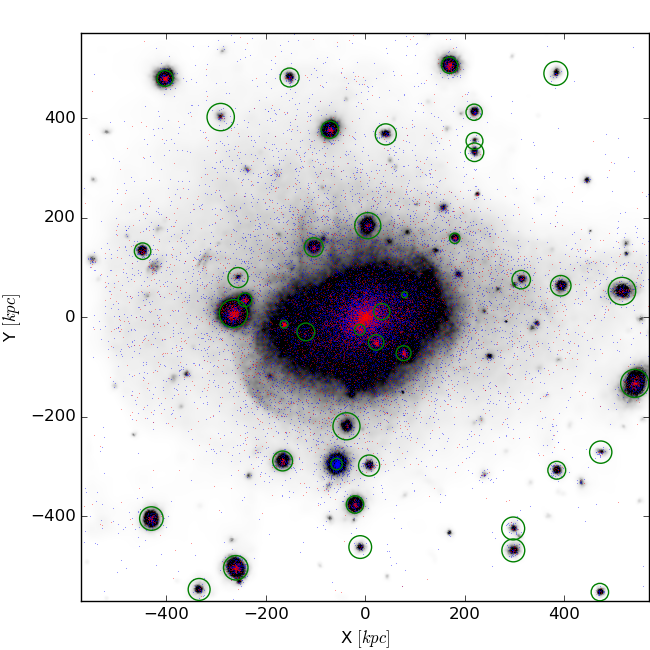}
   \caption{Zoom-in into our most massive galaxy cluster in the sample. 
   Gray scale shows the stellar density and red/blue the tagged GCs candidates. 
   GCs are considered associated to a galaxy if they are within $3$ times the stellar
   half mass radius $3r_{h,*}$ (green circles). 
   }
 \label{fig:zoom-in}
 \end{figure}

  For simplicity, 
we have chosen an infall relation that is independent of redshift
($a_{\rm inf}$ and $b_{\rm inf}$ in Eq.~\ref{eq:harriszin} 
are the same for all infalling galaxies). Theoretical models
give different predictions regarding a possible evolution
of the $M_{\rm GC}$-$M_{\rm vir}$ relation with redshift, 
from nearly no evolution \citep{kruijssen_globular_2015} to
changes by factor $\sim 10$ \citep{choksi_origins_2019} between $z \sim 3$
and present day. The lesson learned from our model is that such
evolution is not {\it needed} based on available observations, 
besides a calibration that accounts for the average fraction of GCs 
lost to tidal removal by the cluster. 
However, this does not mean that normalization of the $M_{\rm GC}$-$M_{\rm vir}$
relation cannot change with time 
\citep[see discussions in e.g., ][]{el-badry_formation_2019, choksi_origins_2019}.
It only means that, given the fair amount of free parameters in our model, 
we choose to adopt an unevolving infall relation and defer
an exploration of changing this hypothesis to future work and plausible 
observational evidence that may justify better such assumption.

Blue (red) points of figure \ref{fig:fig2} shows the resulting blue (red) GCs mass associated to 
galaxies in sample as a function their halo mass at $z=0$.
 We define a GC as
``associated'' to a galaxy if it lays within three times the stellar
half mass radius of a galaxy, i.e., $r< 3r_{h,*}$. This choice is inspired by the typical values used in observations \citep{peng_acs_2008}. However, replicating the exact procedure done on real galaxies is non-trivial and involves the fitting of parametrized profiles to account for an uniform background of not-associated GCs. Instead, here we adopt a simpler definition that scales the size with that of the central galaxy. We have explicitly checked that our main conclusions are not changed significantly by this choice by varying the criteria of association between $2$-$4 r_{h,*}$.

Blue and red lines in Fig.~\ref{fig:fig2} indicate the observational
relation by \citet{harris_dark_2015}. For comparison, we also show in green the relation reported by
\citet{burkert2019} for a combination of all GCs, and shows that our results are in the ballpark of
those observations as well. Once $M_{\rm GC}$ is known for each of our galaxies, we take the number of
candidate dark matter particles associated to each object, $N_{\rm tag}$, and assign a mass 
$M_{\rm GC,\rm inf}/N_{\rm tag}$ to each tagged particle. In this way, all GC particles with the same 
original host galaxy have the same mass.

With this method, each particle tagged as a GC has usually a mass of $\sim 10^3 M_\odot$, 
far below the typical mass of a observed GCs. This happens because the number of {\it tagged} 
candidate particles that are consistent with the right distribution function of the GCs distribution
is greater than the expected number of real GCs in galaxies. Although this procedure allows us 
to have a better sampling of the phase-space of GCs around each galaxy, there are certain
applications where the number of GCs is important (for instance, specific frequency estimations). 
In those cases, it is not adequate to use all the {\it candidate} GCs to compare to observations, 
but instead we re-sample. This is done by adding back all the GC mass (previously distributed among all
tagged particles) and dividing by the typical GC mass chosen to be 
$m_{\rm GC}=10^5 M_\odot$ \citep{brodie_extragalactic_2006}.
This gives us the number of particles that need to be considered per galaxy as ``realistic" GC numbers.
We then randomly select from all candidate GC particles, the needed number of realistic GCs. 
Fig.~\ref{fig:map} shows as an example the most massive galaxy cluster in our simulations
indicating in gray scale the stellar density and highlighted in magenta the tagged GC candidates.

\begin{figure} 
\includegraphics[width=\columnwidth]{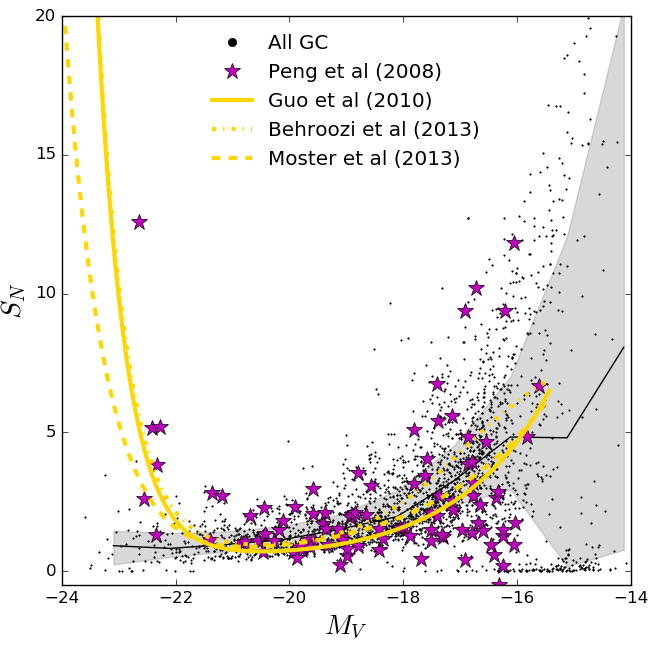}
   \caption{Specific frequency
     ($S_N$) as a function of V-band absolute
    magnitude $M_V$ for all GCs (gray points). The simulation naturally reproduces
    the observed ``U"-shape. Magenta stars are
    observed $S_N$ values in galaxies from the Virgo cluster by \citet{peng_acs_2008}.
    Yellow curves were obtained analytically by assuming the \citet{harris_dark_2015}
    $M_{\rm GC}$ vs $M_{\rm vir}$ power law fit, a constant mass to light ratio
    $M/L_V = 2$ and the stellar mass - halo mass relation described by either 
    \citet{guo_how_2010} (solid) and \citet{moster_galactic_2013} (dashed) or
    \citet{behroozi_average_2013} (dotted).  The solid black curve indicates the median $S_N$ in bins of $M_V$ for our simulated galaxies. The grey area shows the $25\%$-$75\%$ quartiles range. 
}
 \label{fig:s_n}
 \end{figure}


 One of the caveats of our method is that the early destruction of GCs due to
tides, evaporation and infant mortality is not explicitly followed. Instead, our
model is calibrated to reproduce the $z=0$ content of GCs around galaxies, meaning that
those GCs that were disrupted after birth are never part of our sample. This is a necessary compromise, 
since the modeling of such destruction processes can be not only cumbersome, requiring many free parameters
\citep{pfeffer_e-mosaics_2018,el-badry_formation_2019,choksi_formation_2019}, but also requires a numerical resolution that is currently prohibitive for the study of GCs on the scale of galaxy clusters. Consequently, our method is not able to inform us about the early destruction mechanisms of GCs and it has been designed, instead, as a tool to study the predictions of the surviving population at $z=0$ within the cosmological assembly of structures.

\begin{figure} 
\includegraphics[width=\columnwidth]{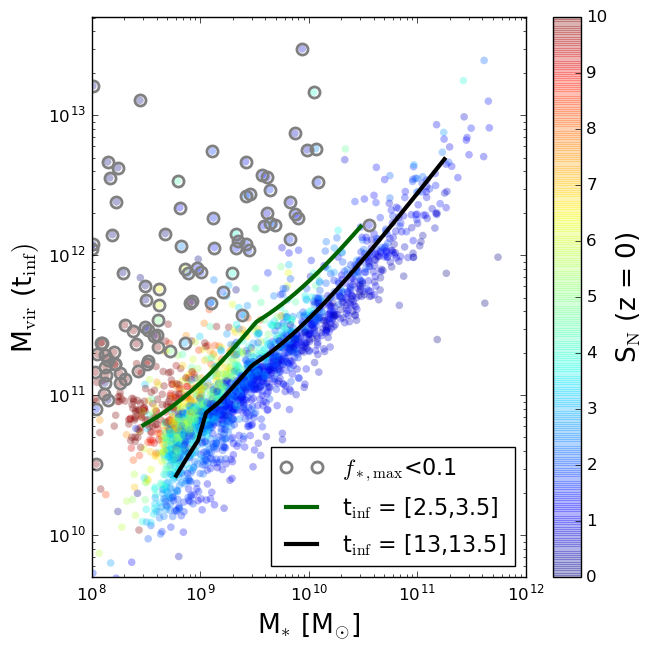}
  \caption{Virial mass at infall as a function of the present day stellar mass in our cluster galaxies. 
  The points are color coded by their $z=0$ specific frequency and
  show that large $S_N$ values are associated to larger
  halo masses at infall. This can be explained as the result of 
  the evolution in the stellar mass-halo mass relation. 
  Black/green lines show the median $M_*$-$M_{\rm vir}$ relation of galaxies 
  with infall times in the ranges $t_{\rm inf}=[13,13.5]$ and $t_{\rm inf}=[2.5$-$3.5]$ Gyr
  respectively. Since the GC mass scales with halo mass (and not $M_*$), 
  that helps explain the larger GC content of galaxies with earlier infall times.
  Additionally, tidal stripping contributes to the scatter. Gray symbols
  highlight objects that lost more than $90\%$ of the stars (see text for details). 
} 
\label{fig:sn_scatter}
\end{figure}

\section{GCs around galaxies in clusters}
\label{sec:s_n}

 After tagging the corresponding GC candidate particles, our technique 
 self-consistently follows the orbits and tidal stripping experienced
 by each galaxy and their population of GCs after they join their
 cluster. As a result of that, the number of GCs that remain
 associated to a given galaxy today may differ from the one that was
 originally assigned. We start by quantifying the distribution and number
 of GCs  around galaxies in our $9$ simulated galaxy clusters in Illustris-1.

Fig.~\ref{fig:zoom-in} shows a zoom-in of the most massive
galaxy cluster within half the virial radius. The background gray
scale map corresponds to the stellar distribution. Superimposed in blue
and red are the tagged {\it candidate} GCs. A close inspection of this
figure indicates that most GCs are found today heavily clustered
around the central and satellite galaxies. As introduce in Sec.~\ref{sec:method}, GCs are considered associated to a galaxy if they lay within $3r_{h,*}$. We indicate this criteria with a green circle. 

We note in Fig.~\ref{fig:zoom-in} that some small
stellar clumps are visible in the grayscale map and have no associated
green circle. These correspond to galaxies that remained below the
minimum mass for tagging ($M_* < 10^8 \; \rm M_\odot$) throughout their
history and will consequently not be considered in the analysis that
follows. In a few exceptional cases a galaxy will be above our minimum
mass criteria but are identified by the merger tree as having been a
satellite of another system at all times. Because the tagging is done
at infall, defined as the last time that a galaxy was a central of its
own group, these objects cannot be tagged and therefore will also be
excluded from the analysis (an example of it can be seen near the
$(-50,-250)$ kpc position in the figure, with a few blue GCs that were
acquired as some smaller companion merged to it). The fraction of such
objects in our sample is below $\sim 2\%$  of the galaxies and
therefore will not strongly impact our results.

\subsection{The GC specific frequency $S_N$}

In observations of galaxy clusters, the specific frequency of GCs
$S_N$, which is defined as the number of GCs per unit $M_V=-15$ of
galaxy luminosity, shows a strong dependence on
galaxy brightness, with a characteristic ``U''-shape indicating large
specific frequencies in bright as well as in dwarf galaxies
\citep[$S_N \geq 5$, ][]{Durrell1996,Lotz2004,peng_acs_2008}. The origin of this behaviour is not well
understood. In particular, in light of the results of field and late type galaxies,
which suggest $S_N \sim 1$-$2$ \citep{harris_globular_1991,Miller1998}. 
Although the large specific frequencies of the
central galaxies may not be surprising given their ability to capture GCs
from orbiting and merged galaxies in the cluster \citep{coenda_tidal_2009}; the
upturn of $S_N$ in the dwarf scale regime is less well understood. 

We use our tagging technique to gain insight into the possible cause
for such trends. Fig.~\ref{fig:s_n} shows $S_N$ as
a function of absolute $V$-band magnitude for our simulated galaxies.  $S_N$ is
computed as,

\begin{equation}
S_N = N_{GC} \times 10^{0.4(M_v + 15)}
\label{eq:s_n}
\end{equation}

\noindent
\citep{harris_globular_1981}, where $N_{GC}$ is the number of
GCs within $r< 3r_{h,*}$ of a galaxy (and assuming an average mass per
cluster $10^5 \; \rm M_\odot$, see
Sec.~\ref{ssec:gcmass}). Encouragingly, the GC specific frequency in
our simulated galaxies, shown in black small dots, follow a similar the familiar
trend found in observations. For comparison we include in
magenta starred symbols a set of observed $S_N$ values for galaxies in
the Virgo cluster from \citet{peng_acs_2008}. Although our simulations tend
to underestimate $S_N$ on the bright end, an effect driven by the
inefficiency of feedback to shut down star formation in high mass
halos for galaxies in Illustris \citep[see Fig. 2 in ][]{genel2014}  artificially lowering our
$S_N$ calculations, the good agreement with intermediate and low mass
dwarfs is a notable success of the GC model.

Our GC tagging method relies mostly on one specific assumption: a single
power-law relation between the mass in GCs and the halo virial
mass at infall. The $S_N$ characteristic ``U''-shape in our model
arises as a consequence of this assumption combined with the well known
non-linear relation between stellar mass (and, consequently,
luminosity) and halo virial mass.  Notice that for the latter we use
the information on stellar mass and luminosity directly from the
Illustris hydrodynamical run and it is not an input of our GC model. 
This conclusion is consistent with previous independent results 
using semi-analytical catalogs or analytic models of GCs formation
\citep[e.g., ][]{peng_acs_2008,harris_catalog_2013,choksi_origins_2019}

To guide the eye, we add in Fig.~\ref{fig:s_n}
yellow curves that are the results of a toy model, as follows. We vary
the halo mass $M_{\rm vir}$ in the range
$\rm log(M_{\rm vir}/M_\odot)=[10, 15]$. For halos in this mass range,
following the $M_{\rm GC}$ - $M_{\rm vir}$ relation from \citet{harris_dark_2015}, we
determine the mass in GCs. As a second step, assuming three different abundance
matching relations $M_*$-$M_{\rm vir}$ \citep[][ with solid, dashed and dotted, respectively]{guo_how_2010, moster_galactic_2013, behroozi_average_2013} we estimate
their $M_*$. Adopting a uniform mass-to-light ratio $\gamma=2$ to go from
stellar mass to $V$-band magnitude and an average GC mass
$M_{\rm GC} \sim 10^5\; \rm M_\odot$, we compute the resulting $S_N$ for
a galaxy with $M_V$ magnitude. This simple model seems to capture the
essence of the simulated and observed trend; thus, it provides a possible
interpretation for the $S_N$ shape.

Our simulations also shed light on the origin on the scatter
of the $S_N$-$M_V$ relation.  The $S_N$ scatter increases 
substantially towards the low mass galaxies in our model, 
in good agreement with observational findings \citep[e.g. ][]{peng_acs_2008, forbes2018}. 
In particular, we note the existence of several dwarf galaxies that
at $z=0$ have no associated GCs. A closer inspection of those objects reveal
that their full GCs content has been tidal stripped by the gravitational 
potential of the host cluster. A more detailed analysis of the scatter
in the $S_N$-$M_V$ relation suggests differences in the infall properties of
surviving galaxies at $z=0$. Fig.~\ref{fig:sn_scatter} shows 
that for a given stellar mass today $M_*$, the corresponding virial
mass {\it at infall} can vary substantially (and therefore, the initial
mass of GCs assigned). As expected, objects with a larger halo mass at
a given present-day stellar mass should display larger specific frequencies
today, as confirmed by the color coding of the points. 

Interestingly, we find that the bulk of the scatter in this plot
is due to an evolving $M_*$-$M_{\rm vir}$ relation with redshift in our
simulations. To illustrate this, we show the median $M_*$-$M_{\rm vir}$ 
relation of those points selected in very narrow ranges of infall times:
$t_{\rm inf}=[2.5,3.5]$ Gyr (green) and $t_{\rm inf}=[13,13.5]$ Gyr (black). 
For halos infalling earlier into the cluster, the same halo mass is populated
by a smaller stellar mass content, resulting in an increased $S_N$ value today.
Such evolution in the $M_*$-$M_{\rm vir}$ relation seen in our simulations 
is also consistent with that determined via abundance matching studies
\citep{moster_galactic_2013,behroozi_average_2013}. 

We have checked that, for those points well above the green curve in 
Fig.~\ref{fig:sn_scatter}, tidal disruption plays a major role. 
Gray circles highlight objects where more than 90\% of the stars 
have been tidally removed.
In most cases tides remove stars and GCs and are conducive to a
small $S_N$ (blue points). However, there is a small subset of heavily
tidally stripped galaxies where the specific frequency increases
to $S_N \geq 8$. These galaxies, located in the dwarf regime
in Fig.~\ref{fig:s_n}, have retained less than 10\% of their stellar
budget but have held on to at least 10 GCs resulting in large specific
frequencies at $z=0$. 
 The distribution of stars follows naturally a shallower profile 
than the cuspy Hernquist profile selected for GCs. Therefore, 
once a galaxy has been significantly depleted of stars, it may enter a 
regime where more stars are lost than GCs, explaining the rise in $S_N$.
However, this might be a direct consequence of
our specific choice of a cuspy radial profile for the tagged GCs
and we prompt the reader to consider $S_N \approx 10$ values and above 
with caution.


Although our assumption of an almost constant slope in the 
$M_{\rm GC}$-$M_{\rm vir}$ relation with redshift partially drives 
some of these results, it is reassuring to see that
estimates of GC formation as a result of mergers might support such an 
assumption \citep[see ][]{choksi_origins_2019}. Moreover, our model makes a
testable prediction: dwarf galaxies with the largest $S_N$ values
within galaxy clusters will correspond to the earliest infalling halos
and therefore should be located preferentially in the inner regions of
clusters. We note that our results are in line with those presented in
\citet{mistani_assembly_2016} with an independent GCs analysis in the Illustris
clusters. \citet{mistani_assembly_2016} also conclude that
earlier infalling dwarfs have larger $S_N$ values, an effect partially
driven by their larger dark matter halo mass --in agreement with our
results-- and that more bursty star formation may increases the
chances of GCs formation. Encouragingly, observational evidence seems
to suggest an increase of $S_N$ for dwarfs near the cluster center
\citep[see Fig. 10 \& 11 in ][]{peng_acs_2008}, in good agreement with our
predictions.

We hasten to add that our prediction for field dwarfs would place them
at $S_N \sim 5$ at $M_V=-16$ (i.e. near the green curves in
Fig.~\ref{fig:s_n}) but they will not scatter upwards. We
defer the comparison to field objects to future work.

\begin{figure} \includegraphics[width=\columnwidth]{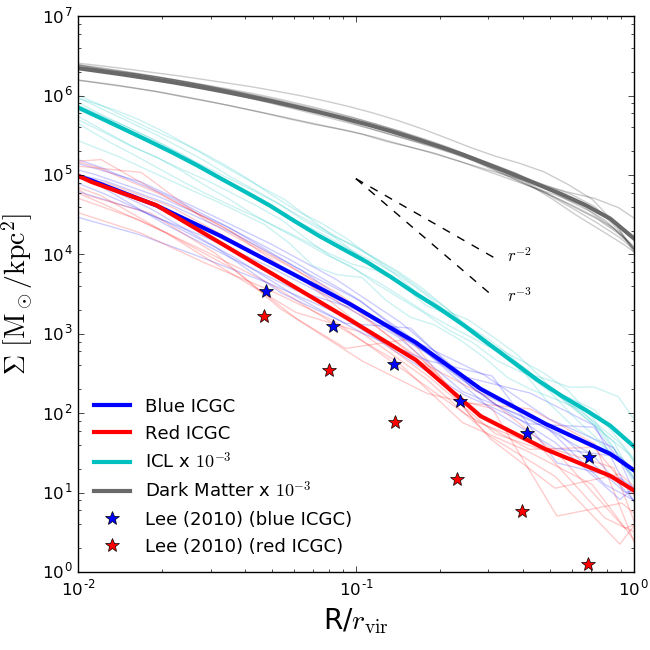}
  \caption{Projected density profiles for the red and blue intra-cluster GCs
  (red and blue curves), using thin lines for the $9$ individual clusters and thick 
  solid for the median in our sample. Encouragingly, the profiles are steeper
  than that of the dark matter in the clusters (gray) and comparable
  to that of the intra-cluster light (stars not associated to galaxies, shown
  in cyan). We compare with measurements in the Virgo cluster from the work
  in \citet{lee_detection_2010}. The simulated red component is shallower
  than detected in Virgo. Normalizations for the dark matter and the intra-cluster
  light are as quoted.} 
  \label{fig:radprof}
\end{figure}

\section{Distribution \& kinematics of intracluster GCs}
\label{sec:rad_vel}

GCs that lay beyond $3r_{h,*}$ from the center of a galaxy (and are
therefore not considered in the $S_N$ calculations of the previous
section) constitute a diffuse and extended intracluster GC (ICGC)
component. They are seemingly not connected to any galaxy today but
are instead linked to the gravitational potential of the host galaxy
cluster. In observations, hints to such intracluster GC population can
be found in early work \citep[e.g. ][]{Forte1982,muzzio_globular_1984,white_globular_1987} 
and confirmed in more recent observations of nearby
galaxy clusters such as Virgo \citep{lee_detection_2010,durrell2014}, Fornax
\citep{bassino2003} and Coma \citep{madrid2018}. We use our simulated
GCs catalog to study this diffuse intracluster component, understand
its origin and determine their potential as dynamical tracers of the
dark matter.

\begin{figure} \includegraphics[width=\columnwidth]{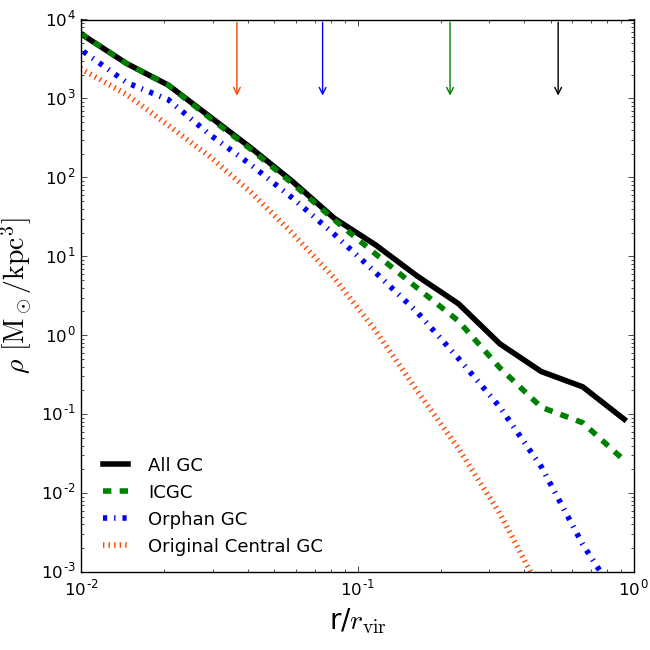}
  \caption{Median (stacked) 3D radial density profiles of simulated GCs 
  separated in different populations: all GCs (solid black curve), intra-cluster GCs (dashed green),
  orphan GCs (defined as those brought into the clusters by a galaxy that
  has been fully tidally disrupted, blue dot-dashed) and GCs originally 
  tagged to the central galaxies in each cluster (dotted red). As shown by
  the half mass radius of each component (matching colors vertical arrow at the
  top of the panel), the intra-cluster component is more centrally concentrated
  than considering all GCs, indicating that the majority of GCs in galaxy
  clusters today are still associated to their host galaxies, in particular, 
  those orbiting in the cluster outskirts.
  } \label{fig:rho} 
  \end{figure}

\subsection{Building the intracluster GCs component}

Fig.~\ref{fig:radprof} shows the projected radial profile of red and
blue intracluster GCs. Thin lines indicate individual simulated galaxy
clusters in Illustris whereas thick solid curves represent the
median. ICGCs have a radial distribution that roughly follows that of
the intracluster light (cyan curves) and that are steeper than the
dark matter (shown in gray), with a surface density profile scaling
$\Sigma_{\rm GC} \sim r^{-2.5}$. A complete census of this diffuse
component in observations is extremely challenging, and as a result
only a few estimates are available to date. For reference, we compare in
Fig.~\ref{fig:radprof} with the SDSS measurement for ICGCs in Virgo by
\citet{lee_detection_2010} shown with starred symbols. 

The blue GCs intracluster component in our simulated clusters seems in
reasonable agreement with \citet{lee_detection_2010} although we note that our
model predicts a shallower profile for the red (or metal rich)
distribution than observed. This is likely due to an insufficient
initial segregation between red and blue components already set at
infall time of individual galaxies in our GCs tagging method. However,
given the non-linearity required for building the ICGC and the
simplicity of our tagging method, the good agreement between the
radial distribution of observed and simulated GCs is  encouraging. To be
conservative, in what follows, we will not distinguish between red and
blue ICGC components and consider it as a single population.

\begin{figure}
  \includegraphics[width=\columnwidth]{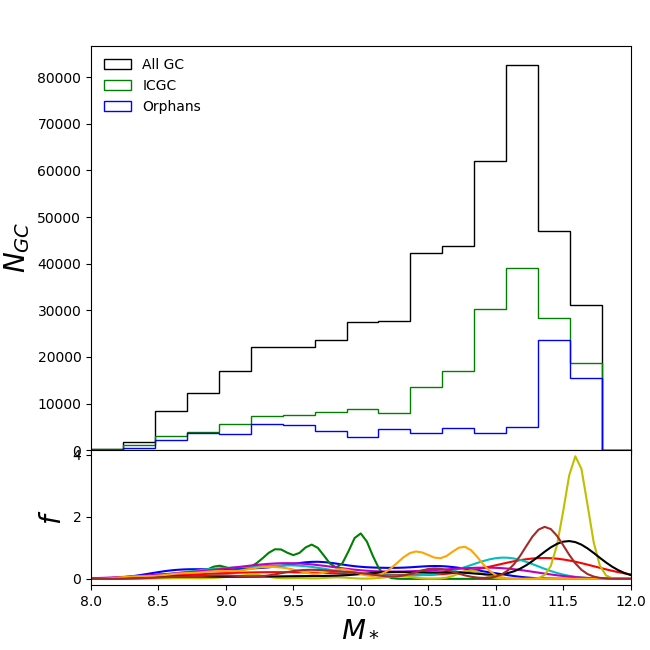}
  \caption{{\it Top:} Distribution of the (stellar) mass of the galaxy progenitors
  that contributed GCs to the total (black) and intra-cluster component (green) in
  all our simulated clusters. 
  These are number weighted distributions, meaning that a galaxy is counted
  as many times as {\it candidate} GCs it contributed, and all simulated clusters
  have been added to a single histogram. The ICGC component is formed
  in its majority from galaxies with stellar mass comparable to
  the Milky Way and above ($M_* \geq 10^{10.5}$). Within the ICGC,
  the orphan population is also dominated by large progenitors (blue curve). However
  this changes from cluster to cluster. {\it Bottom:} distributions 
   (via kernel density estimation) of maximum $M_*$ for galaxies contributing
  GCs to the orphan population in each simulated cluster. 
  Although in most cases orphan GCs were brought in by dwarf-like progenitors
  with $M_* \sim 10^{9.5}\; \rm M_\odot$, in $5$ of our $9$ clusters orphans
  are mostly contributed by major merger events with progenitor galaxies 
  $M_* \sim 10^{11}\; \rm M_\odot$.
 }
  \label{fig:masshist}
\end{figure}

What is the origin of this ICGCs? We find two main contributors to
building the ICGC component, the GCs that were tidally stripped from
(surviving) galaxy satellites in the cluster as well as GCs that were
brought in by galaxies that had fully merged to the central
potential. We will refer to them in what follows as ``orphan'' GCs
since their progenitors not longer exist today. Fig.~\ref{fig:rho}
allows a closer look to this contribution as a function of radius
from the cluster center. We show the 3D median density profile of the
ICGC component in a stacking sample of our simulated
clusters. ICGCs are shown in dashed green and it is split into
contributions from orphans (blue dashed-dotted) and those that were
originally assigned to the central galaxy (dotted red). The remaining
contribution to the ICGCs, accounting for about $62\%$ of this component,
originates from GCs initially associated to galaxies that still survive today 
as satellites in the cluster\footnote{The exact contribution of surviving satellites 
to the ICGC component is only weakly dependent on the adopted cutoff of $3r_{h,*}$ 
to consider a GC as not associated to a galaxy. For example, this changes to 
$54\%$ if the cutoff is extended to $5r_{h,*}$ instead}. Vertical
arrows indicate the half number radius split by components.

GCs initially associated to the central galaxy dominate only in the
inner region and their contribution is not significant beyond
$r/r_{\rm vir} \geq 0.1$. From the rest of the intracluster component,
orphans are more centrally concentrated compared to those from
surviving galaxies. This is consistent with orphan GCs having having had
their progenitor galaxies totally disrupted, which explains their
segregation towards the center of the cluster. For completeness, we
show in Fig.~\ref{fig:rho} the total 3D median profile of GCs
(black). The difference between ICGCs and the total curve indicates,
as expected, that in the outskirts of the galaxy clusters most GCs are
still associated to their host galaxies.

\begin{figure} \includegraphics[width=\columnwidth]{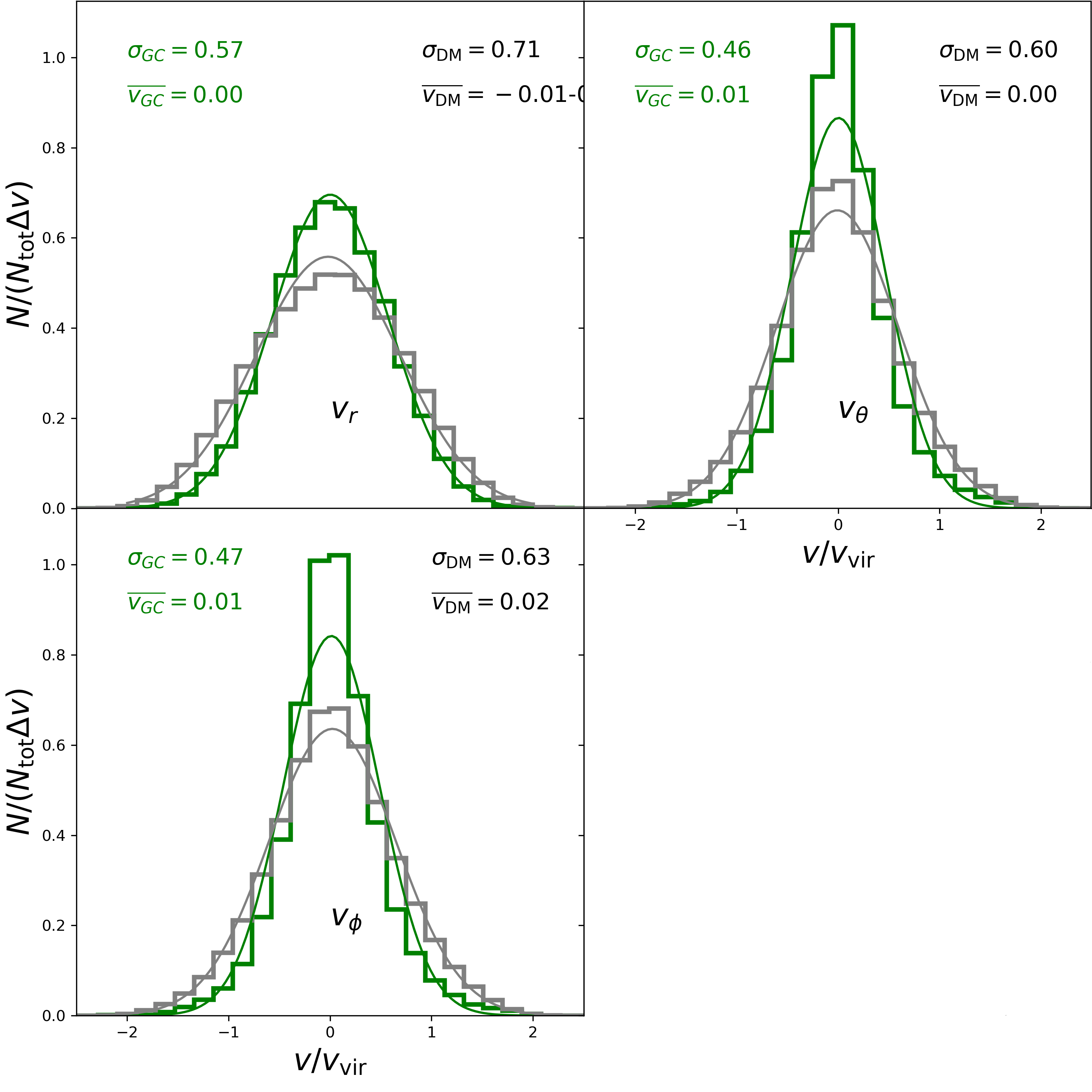}
\caption{Distribution of the spherical velocity components $V_r$, 
$V_\theta$, and $V_\phi$ for the stacked sample of GCs (green) compared
to that of the dark matter in the cluster halos (gray). In general GCs
are good tracers of the underlying dark matter distributions, with velocities
well described by a Gaussian function. We quantify the similarity of the
distributions by quoting the velocity dispersion in each component for
the GCs and for the dark matter (left/right labels, respectively). 
Note also the larger dispersion in the radial direction compared to
the other two tangential components.
} \label{fig:sigmas}
\end{figure}

Because the chemical composition of GCs is linked to that of their
progenitor galaxy \citep[e.g., ][]{Cote1998,Liu2016,Pastorello2015}, 
it is of interest to identify the spectrum
of galaxy masses contributing GCs within a galaxy
cluster. The top panel in Fig.~\ref{fig:masshist} shows the 
added distribution of progenitor galaxy masses for GCs in 
all our simulated clusters, divided according to the different populations. 
Taken as a whole, most GCs today in clusters are linked to massive galaxies with
$M_* \sim 10^{10.5}\; \rm M_\odot$ and above (black solid curve), 
which seems to also apply for the intracluster component (green line). 
We have checked that this is the case when we look individually to each of the simulated clusters. 

On the other hand, the orphan population seems quite diverse.
Although adding all clusters hints to a highly skewed 
distribution towards massive progenitors for the orphan population
(blue line, top panel), when analyzed separately, each cluster
may show a different behavior. We demonstrate this in the bottom panel of
Fig.~\ref{fig:masshist}, which shows for each cluster the progenitor
stellar mass distribution for their orphan population. Because the number
of orphans may vary from object to object, each histogram
has been normalized such that the bins add up to $1$ to ease the comparison. 
We find that about half of the galaxy clusters are heavily dominated by
orphan GCs brought in by one or two very massive $M_* \geq 10^{11}\; \rm M_\odot$
progenitors (major merger events). However, for the other half of the clusters, 
the contribution to the orphan GCs population is dominated instead by 
dwarf-like objects, with median masses $M_* \sim 10^{9.5}\; \rm M_\odot$. 
Such variations should be imprinted in the chemical signatures of 
the ICGC component, providing clues to unravel the past accretion
histories of their cluster hosts. 

Therefore, our simulations indicate an interestingly broad distribution of galaxy masses 
that contributed to the build up of the ICGC component. 
As such, a heterogeneous range of metallicities and
ages are expected to be found in observations, albeit modulated by the radial
trends shown in Fig.~\ref{fig:rho}. We note that histograms in
the top panel of Fig.~\ref{fig:masshist} are number-weighted, meaning that each
candidate GC counts individually and the same galaxy progenitor is
counted as many times as GCs it contributed. For GCs still associated
to galaxies (contributing to the {\it all} curve in black), we
consider the present-day mass of their host galaxy. In the case of
orphans GCs, we count the mass of the galaxy progenitor at their
infall time, since they have totally merged to the cluster by $z=0$ by
definition.

\subsection{Intracluster GCs as kinematic tracers of the dark matter}

\begin{figure} \includegraphics[width=\columnwidth]{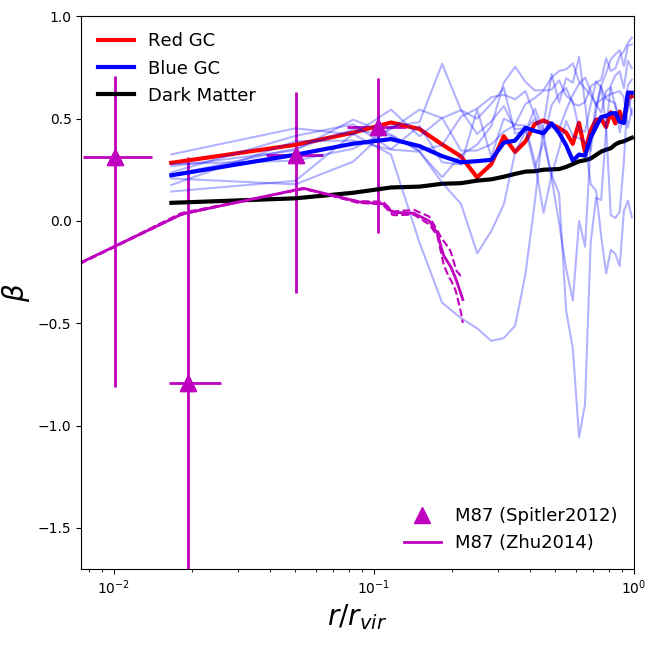}
\caption{Predicted anisotropy profile $\beta$ for the stacked sample of 
red and blue simulated intra-cluster GCs (thick solid red, blue lines, respectively). 
We predict a very radial ($\beta \sim 0.5$) 
orbital structure that is 
comparable to that of the dark matter component in the host cluster 
halo (solid black line).  
Thick lines correspond to the median of the stacked sample of our galaxy clusters and
reproduces  observational in inner regions of M87 \citep{zhu2014}. However, tangentially biased
orbits as measured in the outskirts
of M87 \citep{zhu2014} are more difficult to obtain. The tension may be alleviated
by considering the effects of substructure. Thin solid blue lines show
the intra-cluster GCs for each {\it individual} cluster in our simulation, where
we have correlated the ``dips" in $\beta$ to the presence of infalling groups/substructure.
Notice that the odd external orbital distribution inferred for M87 traces quite closely 
one of our clusters.}
 \label{fig:beta}
\end{figure}

This population of free floating GCs provides also a unique opportunity
to trace kinematically the gravitational potential of the galaxy
cluster in regions where stars or other luminous tracers are scarce.
Obtaining spectroscopic data of several dozens to hundreds of GCs in
the intra-cluster medium of nearby clusters like Virgo or Fornax
is, although expensive, within current to near-future capabilities of
observational campaigns (e.g., Subaru/PFS, Maunakea Spectroscopic Explorer). 
In the MW itself, 3D information of GCs have been used to estimate the enclosed
dark matter mass out to $\sim 100$ kpc, well beyond the stellar and
gaseous disk \citep{watkins2019a}. However, for systems outside our own Galaxy, the
kinematical information will be projected. And how well GCs are
expected to trace the gravitational potential will then depend on
relative biases between dark matter and GCs as well as the orbital
structure predicted for this ICGC component.

Fig.~\ref{fig:sigmas} compares the global kinematics of the ICGCs
(green) to that of the underlying dark matter halo (black) for our $9$
simulated clusters. We show the three spherical components of the
velocity in a spherical coordinates system centered at each cluster. For
this, we have stacked the information of all individual clusters by
scaling the distances and velocities by their respective virial
quantities. The results from individual clusters are not different
from the overall ensemble. We find that the velocity dispersion for
the ICGCs in all three directions is comparable to that of the dark
matter, albeit systematically smaller by $\sim 20$-$25\%$. The more
concentrated radial distribution found for GCs compared to the dark
matter in Fig.~\ref{fig:radprof} is a possible explanation for this
difference. Values for the r.m.s dispersion in all coordinates are quoted individually in each panel for both components to ease the comparison. In all cases, the distributions are well described by a Gaussian function with the measured r.m.s dispersion (indicated by the thin lines).

The radial velocity component for the ICGC as well as for the dark
matter exceeds that of the $\theta$ and $\phi$ components, suggesting
the prominence of radial orbits over tangential ones. This is shown in
more detail in Fig.~\ref{fig:beta} by means of the orbital anisotropy
parameter $\beta = 1 - 2\sigma_r^2/\sigma_t^2$, where 
$\sigma_t^2 = \sigma_\phi^2 + \sigma_\theta^2$ \citep{binney_galactic_1987}. 
Thick solid lines show the median $\beta$ profiles for the ICGC 
separated in red and blue component (although no significant 
difference is seen between both populations). 

We find that the orbits of GCs are expected to be highly radially
biased ($\beta \geq 0.3$) at all radii within the cluster. 
This agrees with the accreted nature of the intracluster population, where galaxies in radial orbits are preferentially tidally disrupted, thus donating their GCs to the cluster potential. 
Dark matter halos in
$\Lambda$CDM galaxy clusters have long been known to be radially 
biased \citep{wojtak2009}.
Fig.~\ref{fig:beta} also shows the median $\beta$ profile of the host
dark matter halos (black curve) and confirms that the ICGC component
is comparably and even slightly more radially biased than the dark
matter in the clusters. Large $\beta$ values therefore seem an
unavoidable prediction for any scenario where structure grows
hierarchically and tidal stripping plays a mayor role
\citep{diemand2005, creasey_globular_2019}. 

A few observational estimates of $\beta$ are currently available in
the literature and offer important constraints to theoretical models
of GC assembly. 
In particular, the kurtosis of the line
of sight velocity dispersion in combination with an estimate of the
circular velocity can be used as an indication of the
orbital anisotropy of a luminous tracer (stars, PNe, GCs) in extragalactic objects
\citep{gerhard1993,vdmarel1993,napolitano2009}. 
Such measurements are particularly challenging due to projection effects and the availability of a discrete number of GCs.

Applying the kurtosis technique to GCs around M87 suggests
that their orbits are radially biased ($\beta > 0$), consistent with our predictions (magenta triangles in
Fig.~\ref{fig:beta} compared to solid lines).  
Interestingly, an
independent analysis of the GCs around M87 presented in
\citet{zhu2014} argues for a $\beta<0$ beyond $r \geq 100$ kpc.
Such preferentially tangential orbits for GCs are difficult to
reconcile with the accretion origin for the ICGC component explored
here \citep[see ][ for a similar conclusion]{creasey_globular_2019}.

As discussed in \citet{spitler2012}, a possible origin for the
negative anisotropy values would be the
preferential tidal disruption of GCs in radial orbits, which would
create an orbital distribution that favors preferentially tangential
orbits. Our method is not well suited to test such hypothesis since
our model is calibrated to reproduce the {\it final} GC content in
halos and the destruction of individual GCs is not explicitly
modeled. Instead, our simulations suggest an alternative explanation
for negative $\beta$ values.  

Thin blue lines in Fig.~\ref{fig:beta}
show the individual anisotropy profile for each of our simulated
clusters and confirm that local deviations toward lower $\beta$ values
are not uncommon. 
We have associated such dips in the profiles with the presence of infalling galaxies that have not yet virialized into the cluster. The GCs of these infalling objects mostly trace the substructure of the cluster (and hence the tangential motion), and they are not directly tracing the cluster potential.


The substructure explanation suggested by our simulations seems well aligned with GCs on possible
tangential orbits in M87, as the Virgo cluster is known to be not
fully relaxed. Moreover, substructure in the kinematics of planetary
nebulae in the Virgo cluster have already been identified \citep{longobardi2018a}.
Our results indicate that {\it localized}
tangentially biased orbits are not inconsistent with the accretion
build up of the ICGC component. However, such $\beta<0$ values should
be local, not extending for more than a few hundred kiloparsecs, offering
a path to proving or disproving such explanation. 

\section{Conclusions}
\label{sec:concl}

We present a novel method to follow the evolution of GCs within galaxy
clusters in cosmological simulations. The method relies on the
observational single power-law relation between mass in GCs and halo
mass to determine the number of GCs associated to a galaxy. By
following the cluster assembly, we trace back all galaxies at infall
time, and by means of a particle tagging technique, we are able to follow the
posterior dynamical evolution of its associated GCs within the
cluster potential. We apply this technique to $9$ of the most massive cluster
halos with $M_{\rm vir}\sim 10^{14}\; \rm M_\odot$ in the cosmological
hydrodynamical simulation Illustris, providing useful constraints to
observations of GCs in systems such as the Virgo cluster.

The successes of this model include reproducing $(i)$ the large
scatter in the specific frequency of GCs, $S_N$ for dwarf galaxies, helping build
the observed ``U''-shape in the $S_N$- stellar mass or magnitude relation;
and $(ii)$ the build up of an intracluster population of
GCs that is spatially distributed following a steeper profile than the
dark matter, in agreement with observations. We find that $S_N \sim 5$
values are naturally expected for dwarf galaxies with $M_V > -18$ as a
result of the non-linear relation between halo mass and stellar
mass. Furthermore, dwarfs that scatter upwards in the $S_N$-$M_V$
relation correspond to the early accreting subhalos onto the cluster,
and their large specific frequency stands from the lower galaxy
formation efficiency at fixed halo mass expected at higher
redshifts. This result is in line with observational evidence of large
$S_N$ values distributed preferentially near the center of the Virgo
cluster \citep{peng_acs_2008}.

We use our simulations to
make predictions about the galaxy progenitors that are expected to be
the main contributors to the intracluster population of GCs. We find
that in numbers the ICGC population is dominated by galaxies
like the MW and above, with masses $M_* \sim 10^{10.5}\; \rm M_\odot$
contributing about $80\%$ of the GCs in this extended and diffuse
component. Furthermore, we find that the tidal origin for the ICGC
translate, on average, into a very radial ($\beta \sim 0.5$) orbital 
structure for the GCs. However, the anisotropy profile can present localized deviations
towards tangentially biased orbits as a result of surviving
substructure of galaxies within the otherwise virialized cluster. It
would be interesting to explore this possibility in light of the
tangentially biased orbital structure measured by the galaxy group
centered at NGC 1407 \citep{spitler2012} and for M87 according to
\citet{zhu2014}.

 An important caveat of our analysis is that the GC content and distribution
for each galaxy is not computed {\it ab-initio}, but instead parametrized
based on empirical relations following Eq.~\ref{eq:rho} and \ref{eq:harriszin}. 
This means that only surviving GCs are tagged, providing no valuable insights on
the formation, dynamics and destruction mechanisms of GCs at birth.
Fully self-consistent models of GC formation and their evolution
within the local interstellar medium embedded within the cosmological
context are the ultimate goal to understand the GC connection to halos
and galaxies. However, the wide dynamical range needed to be resolved
(from sub-pc to several mega-pc scales) are outside the reach of current
numerical simulation capabilities. Important efforts to follow the
formation of stellar clusters are starting to shed light on such
connections, but they are limited to the scale of individual galaxies
\citep{pfeffer_e-mosaics_2018,li_star_2017,renaud_origin_2017,carlberg_globular_2018,Kruijssen2019}, 
higher redshifts \citep{kim_formation_2018} or idealized galaxy mergers set-ups
\citep{karl_disruption_2011}.

We argue that tagging techniques like the one presented here, when
coupled to fully hydrodynamical simulations that follow the formation
of galaxies within $\Lambda$CDM such as Illustris, offer a unique
opportunity to make predictions on the properties of GCs on the scales
of galaxy groups and clusters  until more sophisticated models become
available. This approach is complementary to the
aforementioned theoretical strives on smaller scales and can
potentially be informed by them if more complex modeling is required.
Our current predictions cover the need for a theoretical framework to
interpret the wealth of existing observational data of GCs in nearby
groups and clusters such as Virgo \citep{jordan_acs_2009,durrell_next_2014}, 
Coma \citep{madrid_wide-field_2018}, Fornax \citep{liu2019} and CenA 
\citep{taylor_survey_2017} among others.

\section*{Acknowledgements}
We would like to also thank
the anonymous referee for a detailed and constructive referee report 
that helped clarify and improve the first version of this manuscript.
Authors are grateful for insightful comments and discussions with
Oleg Gnedin, Thorsten Naab, Hui Li and Aaron Romanowski. 
LVS acknowledges support from NASA through the HST Program AR-14583 and 
from the Hellman Foundation. FRA, MGA and HM acknowledge financial support from 
Secyt-UNC and Conicet, Argentina. EWP acknowledges support from the 
National Natural Science Foundation of China under Grant No. 11573002.




\bibliographystyle{mnras}
\bibliography{referencesss.bib} 








\bsp	
\label{lastpage}
\end{document}